\documentclass[usenatbib,preprint]{mn2e}
\usepackage{graphicx}
\usepackage {natbib,aas_macros}
\usepackage {bm}
\usepackage {amsmath,amssymb}
\usepackage {setspace}

\newcommand{\cs}{c_{\rm s}}

\newcommand{\OK}{\Omega_{\rm K}}
\newcommand{\kB}{k_{\rm B}}
\newcommand{\mH}{m_{\rm H}}
\newcommand{\deldel}[2]{\frac{\partial #1}{\partial #2}}
\title[Accreting beyond the $\Omega\Gamma$-limit]
{Primordial protostars accreting beyond the $\Omega\Gamma$-limit:\\
radiation effect around the star-disk boundary}

\author[Takahashi and Omukai]{Sanemichi Z. Takahashi$^{1,2,3}$, Kazuyuki Omukai$^{1}$ \\
$^1$Astronomical Institute, Tohoku University, Aoba, Sendai 980-8578, Japan\\
$^2$Department of Applied Physics, Kogakuin University, Nakano, Hachioji, Tokyo 192-0015, Japan\\
$^3$National Astronomical Observatory of Japan, Osawa, Mitaka, Tokyo 181-8588, Japan
}

\begin{document}
\maketitle
\begin{abstract}
We consider whether the maximum mass of first stars is imposed by the protostellar spin, 
i.e., by the so-called $\Omega\Gamma$-limit, 
which requires the sum of the radiation and centrifugal forces at the stellar surface
be smaller than the inward pull of the gravity. 
Once the accreting protostar reaches such a marginal state, the star cannot spin up more and 
is not allowed to accrete more gas with inward angular momentum flux.
So far, however, the effect of stellar radiation on the structure of 
the accretion disk has not been properly taken into account 
in discussing the effect of $\Omega\Gamma$-limit on the first star formation.
Here, we obtain a series of the steady accretion-disk solutions considering such effect and find solutions 
without net angular momentum influx to the stars with arbitrary rotation rates, in addition to those with finite angular momentum flows.
The accretion of positive angular momentum flows pushes the star beyond the $\Omega\Gamma$-limit, 
which is allowed only with the external pressure provided by the circumstellar disk.  
On the other hand, the accretion with no net angular momentum influx does not result in the spin-up of the star.
Thus, the existence of the solution with no net angular momentum influx 
indicates that the protostars can keep growing in mass by accretion 
even after they reach the $\Omega\Gamma$-limit.
\end{abstract}
\begin{keywords}
stars: Population III -- stars: rotation -- accretion, accretion discs 
\end{keywords}
\section{Introduction}

The first stars are believed to have played crucial roles in the build-up of 
the subsequent structures in the universe \cite[e.g.][]{2009Natur.459...49B}.
The ultraviolet (UV) radiation from them might have contributed in reionizing the intergalactic gas
\cite[][]{1997ApJ...483...21H,2000ApJ...528L..65T,2001ApJ...552..464B, 2002A&A...382...28S}.
Nucleosyntheses inside them and their supernovae enriched the ambient gases with the heavy elements 
\cite[][]{2002ApJ...565..385U,2002ApJ...567..532H, 2013RvMP...85..809K}.
Their remnants also can be seeds of the supermassive black holes observed at $z\gtrsim 6$
\cite[e.g.][]{2010A&ARv..18..279V, 2013ASSL..396..293H}. 
The mass of the first stars mostly determines the extent of their roles in those processes.  

Thanks to a number of studies in the last couple of decades, 
the following picture on the formation of the first stars has emerged. 
The first stars form in minihalos with mass of $\sim 10^6 M_{\odot}$  
\cite[]{1986MNRAS.221...53C,1997ApJ...474....1T,2000ApJ...540...39A}.
The parental gas cloud in such a halo fragments and 
collapses gravitationally due to the H$_2$ cooling \cite[]{2002ApJ...564...23B, 2002Sci...295...93A}. 
A small protostar forms at the center \cite[]{1998ApJ...508..141O,2008Sci...321..669Y}.
After its formation, the protostar grows in mass via accretion of the ambient gas 
\cite[]{2001ApJ...561L..55O,2003ApJ...589..677O}.
With increasing mass of the star, radiation from the protostar becomes so intense and
eventually terminates the accretion, setting the final stellar mass at this moment 
\cite[]{2008ApJ...681..771M,2011Sci...334.1250H}.
Considering a variety of environments in minihalos, their typical mass is found to span 
from dozens to hundreds of solar masses \cite[]{2014ApJ...781...60H,2015MNRAS.448..568H,2014ApJ...792...32S,2016MNRAS.462.1307S}. 
Low-mass stars are also formed as satellites of the massive central star 
by fragmentation of the circumstellar disk 
\cite[]{2008ApJ...677..813M,2010MNRAS.403...45S,2011Sci...331.1040C}.
Some of them are expected to be kicked out from the system by multi-body interaction 
and become free-floating low-mass primordial stars.
The disk fragmentation also causes intermittent accretion onto the main star, 
which would affect the final stellar mass 
\cite[e.g.][]{2012MNRAS.424..457S,2016ApJ...824..119H, 2016MNRAS.462.1307S}. 
Further studies on those processes are required for more quantitative predictions. 

Recently, \cite{2016ApJ...820..135L} proposed that the accretion growth 
of the first stars is regulated by the effect of the protostellar spin. 
In order to maintain the equilibrium configuration of the star, 
the sum of the outward centrifugal and radiation forces cannot exceed 
the inward pull of the gravity at the surface. 
This constraint is called the $\Omega\Gamma$-limit \cite[]{2000A&A...361..159M}.
At this limit \cite[]{1997ASPC..120...83L,1998A&A...329..551L}:
\begin{equation}
 \frac{GM_*}{R_*^2} = \frac{\kappa L_*}{4\pi R_*^2 c}+\Omega_*^2 R_*,
  \label{eq:OmegaGamma-limit1}
\end{equation}
where we have neglected the deformation of the star due to rotation, 
$M_*,\ L_*, \ \Omega_*,$ and $R_*$ are the mass, luminosity, angular velocity, and radius of the protostar, respectively, 
$\kappa$ is the opacity at the surface, $c$ is the speed of light, and $G$ is the gravitational constant.
Using the Eddington ratio $\Gamma\equiv L_*/L_{\rm Edd}$, 
where $L_{\rm Edd}\equiv 4\pi GM_*c /\kappa$ is the Eddington luminosity, 
we can rewrite Equation (\ref{eq:OmegaGamma-limit1}) as:
\begin{equation}
 \left(\frac{\Omega_*}{\Omega_{\rm K*}}\right)^2 = 1-\Gamma,
  \label{eq:OmegaGamma-limit2}
\end{equation}
where $\Omega_{\rm K*}=\sqrt{GM_*/R_*^3}$ is the Keplerian angular velocity at the stellar surface.
During the protostellar accretion phase, the rotation rate 
$\Omega_*/\Omega_{\rm K*}$ increases 
as the accreting gas brings the angular momentum 
into the star \cite[]{2011MNRAS.413..543S}, 
while $\Gamma$ also increases with the mass of the protostar \cite[]{2003ApJ...589..677O}.
As a result, the protostar reaches the $\Omega\Gamma$-limit at some moment. 
The accreting gas can no longer bring the angular momentum in to the star beyond this point.

For further accretion, the angular momentum of the accreting matter 
must be extracted efficiently so that the accretion does not 
increase the specific angular momentum of the star.  
\cite{1991ApJ...370..604P} (hereafter PN91) have found a relation between the rotation 
rate of the protostar and the angular momentum transfer rate from the disk 
to the star by constructing steady accretion solutions 
for the disk and protostellar surface \cite[see also][]{1991ApJ...370..597P}.
According to this, the star spins up by accretion when its angular velocity $\Omega_*$
is smaller than the equilibrium rate of $0.9 \Omega_{\rm K*}$, and vice versa, 
which suggests that the protostar accretes matter from the disk with keeping its spin
at this equilibrium rate.

By applying this result, 
\cite{2016ApJ...820..135L} have discussed the maximum mass of the first stars 
imposed by the $\Omega\Gamma$-limit.
The protostars rotating at the equilibrium rate $\Omega_*/\Omega_{\rm K*}\simeq 0.9$ 
reaches the limit 
when the Eddington ratio becomes $\simeq 0.2$ (by Equation \ref{eq:OmegaGamma-limit2}).
For example, at a typical mass accretion rate 
${\dot M} = 4\times 10^{-3}\  M_\odot {\rm\ yr^{-1}}$, 
$\Gamma\simeq 0.2$ at the mass of $M_*\simeq 7M_{\odot}$ 
according to their stellar evolution calculation.
Further accretion at the same rate would make $\Gamma$ exceed 0.2 
and would cause disruption of the protostar. 
\cite{2016ApJ...820..135L} speculated that the accretion continues beyond $7M_{\odot}$ 
at a reduced rate of $\simeq 10^{-4}\ {\rm M_\odot\  yr^{-1}}$ keeping $\Gamma$ below 0.2.
When the protostar becomes as massive as $\sim 20 M_{\rm \odot}$, its UV luminosity reaches 
$10^{48}\ {\rm s^{-1}}$.
With such high UV luminosity, the accretion is considered to be terminated by 
photoevaporation of the accretion disk \cite[]{2011Sci...334.1250H}.
For this reason, \cite{2016ApJ...820..135L} have concluded that the mass of the first stars is 
limited to about $20M_{\odot}$. 
\footnote{
Here we ignore the effect of the stellar rotation on its structure for simplicity.
Due to the deformation induced by the rotation, 
the radiative flux in the equatorial direction decreases (so-called the gravity darkening) 
and the critical rotation rate becomes smaller than that given by 
Equation (\ref{eq:OmegaGamma-limit2})\cite[]{1998A&A...339L...5G}.
\cite{2000A&A...361..159M} gave the critical Eddington factor considering the gravity darkening.
Using the result of \cite{2000A&A...361..159M}, \cite{2016ApJ...820..135L} 
concluded that the mass of the protostar is limited to about $40M_{\odot}$.
}

Note, however, that PN91's conclusion that the stellar rotation rate $\Omega_*$ remains 
at the equilibrium value of $\sim 0.9\Omega_{\rm K*}$ is 
obtained by the polytropic-gas disk model without considering 
the effect of irradiation by the central star.
The disk structure would be altered near the $\Omega\Gamma$-limit
due to intense radiation from the protostar.
If the accreting gas is supported to some extent by the radiation force, 
it would take longer time for the gas to be delivered to the star and 
its angular momentum might be extracted efficiently enough 
to avoid the stellar spin-up.

Here, we calculate the structure of steady accretion disks taking into account of 
the stellar radiation effect. 
We demonstrate that a protostar at its $\Omega\Gamma$-limit can continue accreting 
the gas without spinning up more even at a slow rotation rate $\Omega_*/\Omega_{\rm K*}<0.9$.

The paper is organized as follows.
In Section \ref{method}, we present basic equations and method for calculation of 
the steady accretion-disk solutions including the irradiation from the protostar.
We examine the obtained disk structures and 
discuss evolution of the star-disk system 
in Section \ref{result}.
Section \ref{discussion} and \ref{summary} are devoted to discussion and summary. 
More discussion on disk structures with different paramenters is presented in Appendix. 

\section{Method}
\label{method}

In this work, we calculate steady solutions for the disks around the accreting first stars.
We do not impose the equality between the centrifugal and the gravitational forces in advance, 
but instead solve the equation of motion in the radial direction.
Assuming that the disk inner-edge contacts with the stellar surface,
we require the angular velocity of the disk at the inner boundary be equal to that of the star.
We consider the effect of the radiation heating and the radiation force from the star, 
and look for the solutions where the angular momentum is not brought to the star from the disk. 

\subsection{Basic Equations}

We assume that the disk is axisymmetric and geometrically thin.
The equations used here are basically the same as those in \cite{1993ApJ...415L.127P} and \cite{1996ApJ...467..749P}.
\subsubsection{Equations of continuity and motion}

For a given mass accretion rate ${\dot M}$, which is assumed to be radially constant,   
the surface density  $\Sigma$ is given by the equation of continuity: 
\begin{equation}
 \Sigma=-\frac{\dot M}{2\pi r v_r}\label{eq:Sigma},
\end{equation}
where $v_r$ is the radial velocity.

Equation of motion in the radial direction is 
\begin{equation}
 v_r \frac{dv_r}{dr} -\Omega ^2 r = -\frac{1}{\Sigma}\frac{d}{dr}\cs^2\Sigma
  - \Omega_{\rm K}^2 r +\frac{\kappa}{c} F_r,\label{eq:dvdr}
\end{equation}
where $\Omega$ is the angular velocity, $\cs=\sqrt{\kB T/\mu \mH}$ the isothermal sound speed, 
$\kB$ the Boltzmann constant, $\mu$ the mean molecular weight, $\mH$ the mass of the hydrogen, $\OK=\sqrt{GM_*/r^3}$ 
the Keplerian angular velocity, and $F_r$ the radiation flux in the radial direction.
The terms in the equation (\ref{eq:dvdr}) from the left to the right 
correspond to the advection, the centrifugal force, 
the gas pressure, the gravitational force, and the radiation force, respectively.
We use the opacity table of \cite{2005MNRAS.358..614M} for low temperatures $\log(T[{\rm K}])\la 4$ 
and of the Opacity Project \cite[cf.][]{1994MNRAS.266..805S} with the abundance of \cite{1998SSRv...85..161G} 
for higher temperatures.
We connect them continuously in $3.75<\log(T[{\rm K}])<4$.

Equation of motion in the azimuthal direction is the equation for the angular momentum conservation:
\begin{equation}
 {\dot M}\Omega r^2 + 2\pi r \nu \Sigma r^2 \frac{d\Omega}{d r}
  = {\rm const.}
  ={\dot J},
\label{eq:Jconserv}
\end{equation}
where $\nu$ is the kinematic viscosity coefficient and ${\dot J}$ is the angular momentum transport rate.
The first and second terms on the left-hand side are the angular momentum transport 
by the advection and that by the viscosity, respectively.
Note that the disk inner-edge 
directly contacts with the stellar surface and 
${\dot J}$ is a free parameter for the torque acting on the star-disk 
boundary between the star. This differs from the usual 
assumption of the zero torque at the inner boundary, 
which corresponds to ${\dot J}={\dot M}j_{\rm in}$, 
where $j_{\rm in}$ is the specific angular momentum 
$\Omega r^2$ at the inner edge.
We can rewrite Equation (\ref{eq:Jconserv}) as 
\begin{equation}
 \frac{d\Omega}{dr} = \frac{v_r}{\nu}\left(\Omega-\frac{j}{r^2}\right),\label{eq:dOmegadr}
\end{equation}
where  $j \equiv {\dot J}/{\dot M}$.
We use the $\alpha$-viscosity model \cite[]{1973A&A....24..337S},
\begin{equation}
 \nu=\alpha \frac{\cs^2}{\OK}.
\end{equation}
We adopt $\alpha=10^{-2}$ which is a typical value for magnetohydrodynamic turbulence in circumstellar disks \cite[cf.][]{1996ApJ...464..690H,2004ApJ...605..321S}.

\subsubsection{The energy equation}
The energy equation is 
\begin{equation}
 Q_{\rm vis}+Q_{\rm rad} +Q_{\rm adv}+Q_{F_r} +Q_{\rm chem}= 0,
\label{eq:energy}
\end{equation}
where the $Q$'s are the heating rates (negative values mean the cooling) per unit area by 
the viscous heating ($Q_{\rm vis}$), radiation from the surface 
in the vertical direction ($Q_{\rm rad}$), 
chemical reactions ($Q_{\rm chem}$), advection ($Q_{\rm adv}$) and  
the radially diffusing radiation ($Q_{F_r}$), respectively.
Note that all those quantities are functions of radius and Equation (\ref{eq:energy}) 
must be satisfied at all radii.
We obtain the disk structure solving the equations (\ref{eq:dvdr}), (\ref{eq:dOmegadr}), (\ref{eq:dTdr}), and (\ref{eq:energy}) 
numerically with the relaxation method. 

The heating/cooling rate by each processes are given as follows.

The viscous heating rate per unit volume is 
\begin{equation}
 q_{\rm vis} =\nu\rho\left(r\frac{d\Omega}{dr}\right)^2,
\end{equation}
where $\rho=\Sigma/(2H)$ is the mass density \cite[cf.][]{2008bhad.book.....K}. 
Integrating this equation in the vertical direction and substituting the equation (\ref{eq:Sigma}), we obtain the viscous heating rate per unit area 
\begin{equation}
 Q_{\rm vis} 
=-\frac{\nu {\dot M}}{2\pi r v_r}
  \left(r\frac{d\Omega}{dr}\right)^2.
\end{equation}

The radiation cooling in the vertical direction is  
\begin{equation}
 Q_{\rm rad} = -2F_z,
\end{equation}
where $F_z$ is the radiation flux at the disk surface, which is given as follows \cite[cf.][]{1990ApJ...351..632H}:
\begin{equation}
 F_z = \frac{\sigma T^4}
  {\frac{3}{4}(\frac{\tau}{2}+\frac{1}{\sqrt{3}}+\frac{1}{3\tau})},
\label{eq:Fz}
\end{equation}
where $\tau=0.5\kappa\Sigma$ is the optical depth and $\sigma$ is the Stefan-Boltzmann constant.
The right hand side of Equation (\ref{eq:Fz}) is proportional to $\tau^{-1}$ in the optically thick limit and $\tau$ in the optically thin limit, and smoothly connects between these limits.

The energy change due to the advection $Q_{\rm adv}$ is 
\begin{equation}
 Q_{\rm adv} =  -\frac{1}{r}\frac{d}{dr}\left[r v_r(E+\Pi)\right]
  +v_r\frac{d\Pi}{dr}+\int_{-H}^{H}v_z\deldel{p}{z}dz,\label{eq:Qadv1}
\end{equation}
where $\Pi=\cs^2\Sigma+(2/3)aT^4H$ is the sum of the gas and the radiation pressure integrated in the vertical direction. 
The compressional heating is also included in $Q_{\rm adv}$.
Equation (\ref{eq:Qadv1}) can be rewritten as
\begin{equation}
 Q_{\rm adv} = -\xi \frac{v_r \Pi}{r},
\end{equation}
where $\xi$ is given as follows:
\begin{equation}
 \xi = \frac{\gamma+1}{2(\gamma-1)}\frac{r}{T}
  \frac{dT}{dr}
  +\frac{r}{v_r}\frac{dv_r}{dr}
  +\frac{5}{2}.
\end{equation}
See \cite{2008bhad.book.....K} for derivation of this equation.
\footnote{
This expression of $\xi$ is exact only for the vertically isothermal disk in which the radiation pressure is negligible.
The details of the expression of $\xi$, however, does not affect the result because our calculation shows that $Q_{\rm adv}$ is negligible over a wide range of the radius of the disk.
}
We adopt the specific heat $\gamma=5/3$ since the gas is completely ionized around the protostar.
Since the self-gravity of the disk is negligible, the scale height is given by the hydrostatic equilibrium in the vertical direction 
 between the pressure gradient force $P/H$ and the stellar gravitational force $GM_*\rho H/r^3 = \OK^2\rho H$ \cite[cf.][]{1973A&A....24..337S}:
\begin{equation}
 \frac{P}{H} = \OK^2\rho H,
\label{eq:hyd_sta}
\end{equation}
where the total pressure $P$ is the sum of the gas and radiation contributions:
\begin{equation}
 P=\cs^2\rho+\frac{a T^4}{3}.
\label{eq:Ptot}
\end{equation}
From equations (\ref{eq:hyd_sta}) and (\ref{eq:Ptot}), we obtain the scale height
\begin{equation}
 H=\frac{aT^4/3}{\Sigma \OK^2}+\sqrt{\left(\frac{aT^4/3}{\Sigma \OK^2}\right)^2+\frac{\cs^2}{\OK^2}}.\label{eq:H}
\end{equation}

The radial radiation flow strongly affects the energy balance in the inner disk.
Since the disk is optically thick around the star, the radial radiative flux $F_r$ 
is evaluated with the diffusion approximation:
\begin{equation}
 F_r = -\frac{16\sigma T^3}{3\kappa\rho}\frac{dT}{dr}.\label{eq:dTdr}
\end{equation}

The chemical heating rate is given by
\begin{equation}
Q_{\rm chem}=Q_{\rm H_2}+Q_{\rm H}+Q_{\rm He}+Q_{\rm He^{+}},
\end{equation}
where the terms on the right hand side are the rates of energy change due to the dissociation of the hydrogen molecules, 
the ionization of the hydrogen atoms, the first and second ionization of the helium, respectively.
We adopt the helium mass fraction $Y=0.25$ and assume that the abundances are 
in the chemical equilibrium. 
The energy change rate $Q_{\rm H_2}$ due to the H$_2$ dissociation is calculated from the H$_2$ number density $n_{\rm H_2}$:
\begin{equation}
 Q_{\rm H_2} = \chi_{\rm H_2}\frac{1}{r}\frac{\partial}{\partial r}
  (2Hr n_{\rm H_2} v_r),
\end{equation}
where $\chi_{\rm H_2} =4.48$ eV is the binding energy of a hydrogen molecule.
Similarly, $Q_{\rm H}$, $Q_{\rm He}$, are $Q_{\rm He^{+}}$ are given as follows:
\begin{equation}
 Q_{\rm H} = -\chi_{\rm H}\frac{1}{r}\frac{\partial}{\partial r}
  (2Hr n_{\rm H^+} v_r),
\end{equation}
\begin{equation}
 Q_{\rm He} = \chi_{\rm He}\frac{1}{r}\frac{\partial}{\partial r}
  (2Hr n_{\rm He} v_r),
\end{equation}
\begin{equation}
 Q_{\rm He^+} = -\chi_{\rm He^+}\frac{1}{r}\frac{\partial}{\partial r}
  (2Hr n_{\rm He^{++}} v_r),
\end{equation}
where $n_{\rm H^+}$, $n_{\rm He}$, and $n_{\rm He^{++}}$ are the number densities of ${\rm H^+,\ He^+,\ and\ He^{++} }$ respectively, and $\chi_{\rm H} =13.6$ eV, $\chi_{\rm He} =24.6$ eV, and $\chi_{\rm He^+} =54.4$ eV are the binding energies for the ionization of  ${\rm H,\ He,\ and\ He^{+} }$, respectively.

\subsection{Boundary Conditions}
In order to solve the four equations (\ref{eq:dvdr}), (\ref{eq:dOmegadr}), (\ref{eq:dTdr}), and (\ref{eq:energy}), 
we need to supplement four boundary conditions.
According to the previous work \cite[]{1993ApJ...415L.127P}, we adopt the following inner and the outer boundary conditions.

\subsubsection{Inner Boundary Conditions }
Assuming that the inner edge of the disk directly contacts with the stellar surface, 
we take the inner disk radius $r_{\rm in}$ at the star radius $R_*$.
The angular velocity at the inner edge coincides with that of the star,
\begin{equation}
 \Omega(r_{\rm in}) = \Omega_*.
\end{equation}
The incoming radiation flux $F_r$ at the inner edge is provided by that from the star $F_*$:
\begin{equation}
 F_r(r_{\rm in}) = F_*=\sigma T_*^4,\label{eq:Fr}
\end{equation}
where $T_*$ is the stellar effective temperature.

\subsubsection{Outer Boundary Conditions}
We set the outer radius at $ r_{\rm out} = 30 \ {\rm au}$, which is large enough 
for the outer boundary conditions not to affect the disk structure around the star.
At the outer boundary, we assume the Keplerian rotation:
\begin{equation}
 \Omega(r_{\rm out}) = \OK(r_{\rm out}),
\end{equation}
and the thermal equilibrium between the radiative cooling and the viscous heating:
\begin{equation}
 Q_{\rm vis}(r_{\rm out})+Q_{\rm rad}(r_{\rm out})=0.
\end{equation}

\subsection{Protostellar models}
\begin{figure}
  \begin{center}
   \includegraphics[width=80mm]{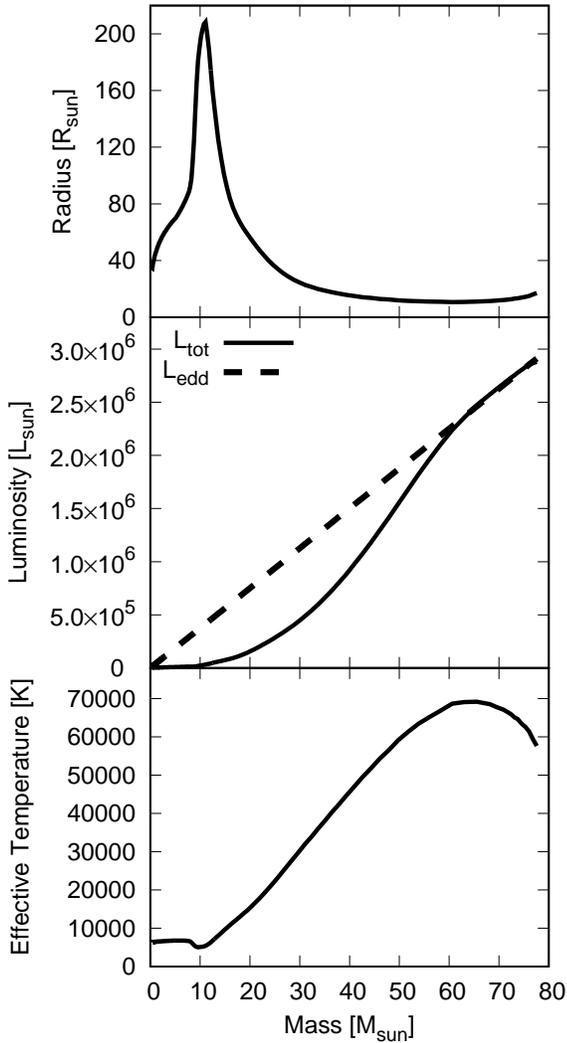}
   \caption{The protostellar evolution model used in this work, which is adapted from 
   Hosokawa et al. (2012).
   We show the radius, the total luminosity (i.e., the sum of the stellar and accretion luminosities)
and effective temperature of the protostar, from top to bottom, as functions of the stellar 
mass. The stellar evolution is calculated with the constant accretion rate 
of ${\dot M}=4\times 10^{-3}\ M_{\odot}\ {\rm yr^{-1}}$ and the shock boundary condition.
}
   \label{fig:prost}
  \end{center}
\end{figure}

\begin{figure}
\begin{center}
  \includegraphics[width =8cm]{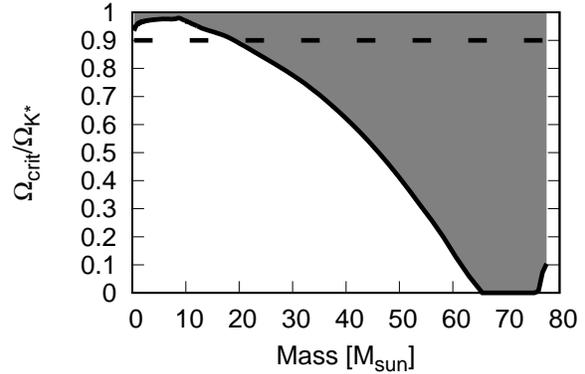}
  \caption{The critical stellar rotation rate at the $\Omega\Gamma$-limit
obtained for the star shown in Figure \ref{fig:prost} normalized by the Keplerian rate at the stellar surface. The shadowed region is prohibited. 
The electron scattering value $\kappa=0.35\ {\rm cm^2\ g^{-1}}$ is used for calculating 
$\Omega_{\rm crit}$ from equation (2).
The dashed line shows the equilibirium angular velocity 
$\Omega_*/\Omega_{\rm K*}=0.9$ obtained by PN91.
}
\label{fig:Omega_crit}
\end{center}
 \end{figure}

We use the result of protostellar evolution calculation at
typical accretion rate of $ {\dot M} = 4\times10^{-3} \ M_{\odot}\ {\rm yr^{-1}}$ 
with the shock boundary condition by \cite{2012ApJ...756...93H}
for the stellar parameters.
Figure \ref{fig:prost} shows the radius, the total luminosity, and the effective temperature as a function of 
the protostellar mass.
In the middle panel, also shown by the dashed line is the Eddington luminosity for the electron scattering opacity.
Using the radius and the luminosity as functions of the stellar mass shown in Figure \ref{fig:prost}, 
we calculate the critical angular velocity $\Omega_{\rm crit}$ where the star reaches 
the $\Omega\Gamma$-limit (Equation \ref{eq:OmegaGamma-limit2}) and show it in Figure \ref{fig:Omega_crit}.
\footnote{
Here, we use the electron scattering opacity $\kappa=0.35\ {\rm cm^2\ g^{-1}}$ 
to evaluate $L_{\rm Edd}$ assuming the complete ionization.
In the actual protostellar evolution calculation used in our model, 
the opacity can be smaller than $0.35\ {\rm cm^2\ g^{-1}}$ around the surface.
From Figure \ref{fig:Omega_crit}, it looks that $\Omega_{\rm crit}=0$, 
i.e. any finite rotation is not allowed by the $\Omega\Gamma$-limit, 
for $M_{\ast} \ga 70 M_{\odot}$ since the total luminosity is larger than the nominal 
$L_{\rm Edd}$ evaluated by the electron scattering opacity 
(see the middle panel of Figure \ref{fig:prost}).
Of course, although very small, some finite rotation rate is allowed if we consider 
the fact that the actual opacity is smaller than 0.35~${\rm cm^2\ g^{-1}}$.
}
As envisaged in PN91, if the protostar evolves with the 
angular velocity $\Omega_*/\Omega_{\rm K*}=0.9$ 
(dashed line in Figure \ref{fig:Omega_crit}), 
the star reaches the $\Omega\Gamma$-limit at the mass of $M_{\ast}\simeq 20M_{\odot}$.

\section{Results}
\label{result}
\subsection{Disk structure at the $\Omega\Gamma$-limit}

By calculating the steady-disk structure irradiated from the protostar, 
we find that the star can accrete the gas without spinning up 
even with rotation rate below the PN91's equilibrium value 
($\Omega_{\ast}/\Omega_{\rm K*}<0.9$)
in the mass range we have studied $M_*=5-70M_{\odot}$.

\begin{figure}
 \includegraphics[width=8.5cm]{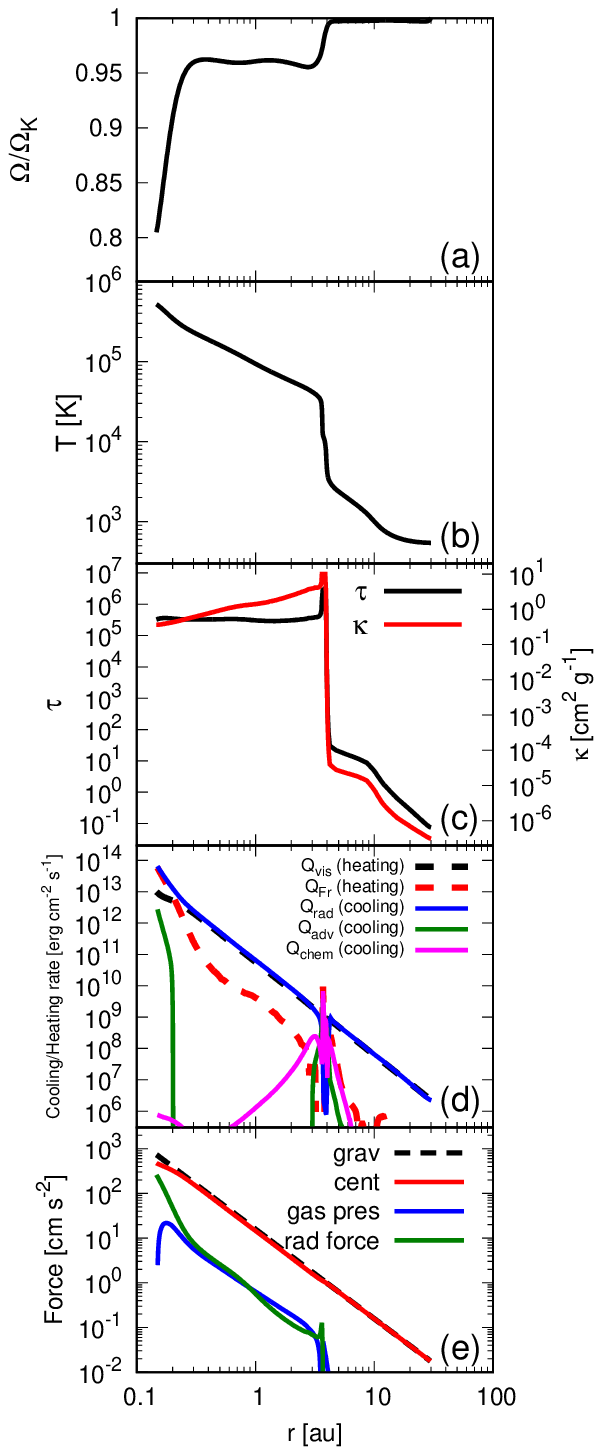}
\caption{
Disk structure with stellar mass $M_*=26M_{\odot}$, 
stellar rotation rate $\Omega_*/\Omega_{\rm K*}\approx0.8$ and no net angular momentum inflow to
the star $j=0$.
The five panels from top to bottom show the angular velocity (panel a), the temperature (b), 
the optical depth and opacity (c),
the heating and the cooling rates per unit area by individual processes (d), 
and the forces per unit mass (e), respectively, as functions of the radius.
In panel (d), heating/cooling rates are shown by the absolute values of $Q$'s.
}
\label{fig:M26}
\end{figure}

As an example, we present the disk structure around a protostar with mass $M_*=26M_{\odot}$
rotating at the critical rate for the $\Omega\Gamma$-limit, 
$\Omega_{\ast}=\Omega_{\rm crit} \simeq 0.8 \Omega_{\rm K*}$.
Figure \ref{fig:M26} shows such solution with $j=0$, i.e., no net angular momentum 
is brought into the star from the disk.
From the $\Omega/\Omega_{\rm K}$ and $T$ distributions (Figure \ref{fig:M26} a and b), 
the disk can be divided into the three regions, i.e., the outer ($\gtrsim 4$ au), 
intermediate (0.3-3 au) and inner ($\la 0.3$ au) regions.  
In the outer region ($r\gtrsim 4$ au), the disk is optically thin and 
efficient radiative cooling keeps
the temperature relatively low at $<$ a few $10^3$ K.
The gas composition is still molecular, and the rotation rate is almost Keplerian 
($\Omega/\OK=1$) with little pressure support.
Temperature increases inward gradually until $T\sim 3000$ K. 
At this point, it increases suddenly by an order of magnitude 
within a thin ($\sim$ 1 au) transition zone
because the opacity and optical depth jump up due to the contribution 
of the $\rm H^{-}$ bound-free absorption and the resultant suppression of radiative cooling (Figure \ref{fig:M26} c and d).
Slightly outside the transition zone, the gas is heated more 
by the diffusive radiation from the inner hot part 
than by the viscous heating. 
This radiative heating balances with the ionization cooling and the advection cooling 
(Figure \ref{fig:M26} d).
\footnote{
Within the transition zone, the opacity increases by a few orders of magnitude.
Such a drastic change of the opacity inhibits us from obtaining convergent solutions 
with the relaxation method.
To avoid this difficulty, we have set an upper floor 
on the opacity to be 10 $\rm cm^2\ g^{-1}$ in looking for the solution.
This causes slight deviation in the disk structure within the transition zone 
from the correct one with the actual opacity. 
This procedure only affects the result in the very thin transition zone.
We do not try to discuss the detailed structure of the transition zone in this work.
}
Further inward from the transition zone is the intermediate region 
($0.3\ {\rm au} \lesssim r \lesssim 3\ {\rm au}$), 
where the temperature is high at several $10^4$K due to 
the large  optical depth ($\tau=$several $10^5$).
The gas is now almost fully ionized and so the opacity is close to the electron scattering value.
In the innermost region, radiation pressure has a non-negligible share in the force balance (Figure \ref{fig:M26} e), 
so that the angular velocity is slightly reduced from the Keplerian $\Omega/\OK\sim 0.95$ 
(Figure \ref{fig:M26} a).
Both in the outer and intermediate regions, the viscous heating almost balances 
with the radiative cooling from the disk surface
except around the very thin transition zone.
In the inner region ($r\lesssim 0.3$ au), the disk structure largely deviates from that of 
the standard disk due to the effect of irradiation from the protostar.
Here, the angular velocity is significantly smaller than the Keplerian value:
at the inner edge, $\Omega/\OK\sim0.8$ (Figure \ref{fig:M26} a) and 
the radiation force reaches $\simeq 40$\% of the gravitational pull 
(Figure \ref{fig:M26} e).
The viscous heating rate is reduced by this small angular velocity and 
the radiative heating by the protostar now balances with the radiative cooling 
from the disk surface (Figure \ref{fig:M26} d).

Unlike the case of PN91, 
the star does not receive net angular momentum from the disk in our case
despite the stellar rotation rate ($\Omega_*/\Omega_{\rm K*}\simeq 0.8$) is smaller 
than the PN91's equilibrium rate ($\simeq 0.9$).
In the inner region of the disk, 
the angular velocity is smaller than that of the star
 ($\Omega \lesssim \Omega_* =0.8\Omega_{\rm K*}$)
and thus the gravitational attraction exceeds the centrifugal force.
Without the outward radiation force, the disk material would fall onto the star within a dynamical timescale.
The accreting gas would deliver its angular momentum to the star 
keeping its original value with little loss 
because the timescale for the angular momentum transfer is much longer than the dynamical timescale. 
The advection of the angular momentum from the disk would make the star to rotate faster and faster up to the Keplerian rate.
In reality, however, the accreting matter is supported not only by the centrifugal force but also the radiation force
against the gravity. 
The disk material now arrives on the star in a timescale long enough for angular momentum to be transferred outward.
The accretion of such low angular momentum matter does not result in the spin-up of the star, i.e., 
$\Omega_*/\Omega_{\rm K*}$ does not increase.
This suggests that the protostar can continue accretion 
even after it reaches the $\Omega\Gamma$-limit. 
We compare our results with that of the previous work in Section \ref{comp} and 
discuss the reason why different conclusions are reached.

\subsection{Temporal evolution of the protostar-disk system}
\label{evolv}
Here we discuss the plausible evolution of 
the protostar and accretion disk system considering the effect of the $\Omega\Gamma$-limit.
Following PN91, 
we speculate that a slowly rotating star spins up, while fast rotator spins down, 
by accretion. That is, the protostellar rotation approaches some equilibrium rate by accretion. 
Suppose a slowly rotating star receives a matter rotating at the Keplerian rate  
$\Omega_{\rm K*}$ and the torque is negligible at the disk inner edge.
The angular momentum influx to the star is ${\dot M}\Omega_{\rm K*}R_*^2$, which 
is consistent with PN91, where it is 
$\simeq {\dot M}\Omega_{\rm K*}R_*^2$ for $\Omega_*/\Omega_{\rm K*}\lesssim 0.9$.
The primordial star acquiring the angular momentum at such rate reaches the $\Omega\Gamma$-limit at 
$M_{\ast}=0.2-0.3M_\odot$ according to \cite{2016ApJ...820..135L}. 
Although disk solutions with inward angular momentum flux still exist even for the stars on the verge of the $\Omega\Gamma$-limit, 
the stars would spin up at a rate higher than the critical value, with its surface confined by the external pressure from the disk (Appendix \ref{Omega_dependence}).
Such a situation, however, would not be able to last permanently.
To push back the stellar surface rotating at ever increasing rate, 
correspondingly large inward pressure force and thus a very massive disk will be required. 
This may be accomplished as the accretion from the surrounding envelope to the disk would continue regardless 
of the stellar $\Omega\Gamma$-limit. 
Another possibility is that, to dump accumulated matter, the disk would adjust itself eventually to the solution with no or little 
net angular momentum inflow as long as such a solution exists.
We here assume the latter is indeed the case, i.e., that the star will keep rotating at the critical rate 
$\Omega_*=\Omega_{\rm crit}$ 
once it reaches the $\Omega\Gamma$-limit, and the accretion continues
with no net angular momentum influx $j= 0$.

\begin{figure}
 \begin{center}
  \includegraphics[width=8cm]{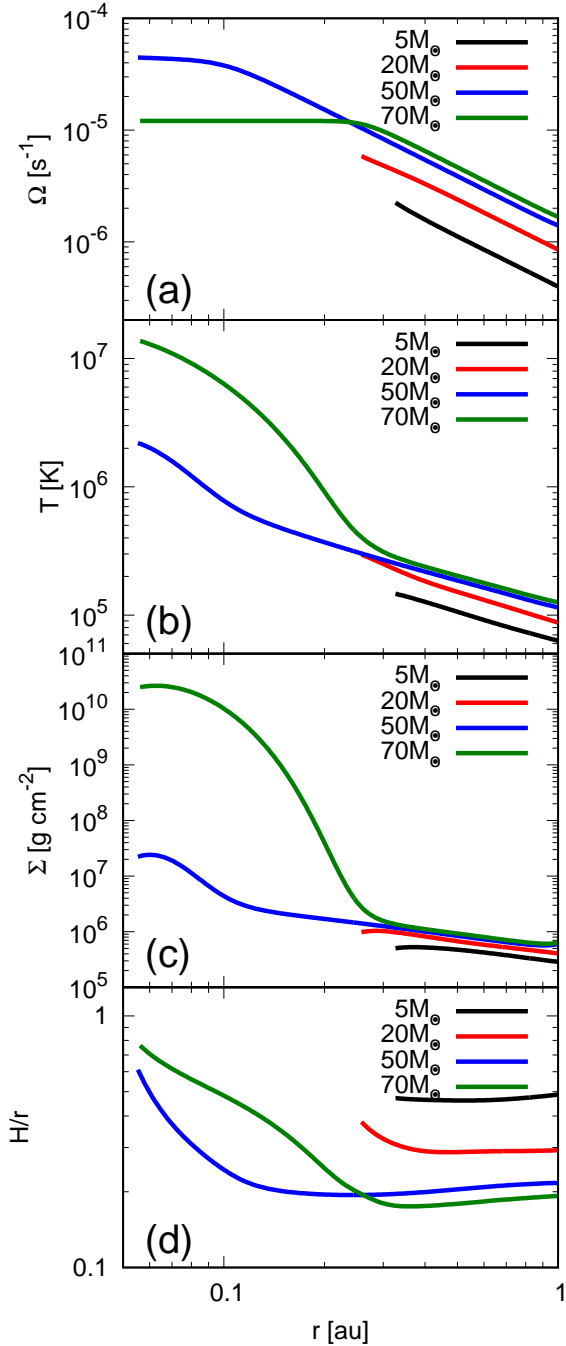}
  \caption{
Disk structures with $\Omega_*=\Omega_{\rm crit}$ and $j=0$ when the stellar mass is $M=5, 20, 
50$, and $70M_\odot$, which correspond to 
$\Omega_{\rm crit}/\Omega_{\rm K*}=$0.93, 0.86, 0.41 and 0.096, respectively.
The four panels from top to bottom show the radial distribution of (a) angular velocity, 
(b) temperature, (c) surface density, and (d) aspect ratio. 
The inner radius of the disk corresponds to the protostellar radius.
}
\label{fig:evolv}
 \end{center}
\end{figure}

In the following, we present the disk structures with $\Omega_*=\Omega_{\rm crit}$ and $j=0$ 
for a series of stellar models with different masses,
which can be regarded as a temporal sequence. 
We defer discussion on other cases to Appendix \ref{Omega_Jdot}.
Figure \ref{fig:evolv} shows the structure of such disks 
at the protostellar mass $M_*=5, 20, 50$, and $70M_\odot$.
Note that the protostar is able to accrete the matter without receiving 
the angular momentum in those solutions in contrast to 
the case of PN91 where it is allowed only for $M\lesssim 20M_{\odot}$. 
As shown in Figure \ref{fig:Omega_crit}, the ratio of the critical to the Keplerian rotation rate $\Omega_{\rm crit}/\Omega_{\rm K*}$ decreases with increasing stellar mass and becomes as low as $\simeq 0.1$ at $70M_{\odot}$. 
In fact, the star can accrete the gas without spinning up even with such a small value of $\Omega_{\rm crit}$.
Although we have calculated the disk solutions only up to $70M_{\odot}$, we expect the solutions with $j=0$ continue to exist for corresponding $\Omega_{\rm crit}$ values.

It is worth noting that the scale height of the inner disk is very large with $H/r\sim O(1)$ 
for high-mass cases with $M_*=50$ and $70M_\odot$.
\footnote{
Since the thin disk approximation breaks down in this region, the accuracy of the calculation is not high.
However, such a thick envelope-like structure is expected to appear anyway.
}
Recall that the disk rotates almost rigidly in the inner edge with 
$\Omega = {\rm const.} \simeq \Omega_{\rm crit}$, while $\Omega \simeq \OK$ in the outer region for those cases.
In the disk with $j=0$, the infall velocity should be small when the shear is small
as $v_r\propto d\Omega/dr$ from Equation \ref{eq:dOmegadr}.
Thus, the infalling gas gets stuck and the surface density becomes large 
around the inner edge (Figure \ref{fig:evolv} c).
The resultant large optical depth makes the radiative cooling inefficient 
and the temperature becomes high in this region.
As a result, the protostar is covered by a hot envelope-like structure once $\Omega_{\rm crit} \ll \Omega_{\rm K*}$.
Note, however, that we have used the stellar model neglecting the back reaction from the disk. 
In reality, this hot envelope would affect the outgoing radiation field (see Section \ref{Teff}).

Note that, due to very high column density in the inner part, the nominal Toomre's $Q$ value can be smaller than unity. 
This, however, does not immediately mean fragmentation occurs in such a region. 
Since the aspect ratio reaches almost unity and the gas is supported by the thermal pressure rather than the centrifugal force, 
it is more like a part of the star than the disk. Thus the Toomre's stability criterion cannot be applied.

\section{Discussion}
\label{discussion}

\subsection{Comparison with previous works}
\label{comp}

Based on PN91's result, 
\cite{2016ApJ...820..135L} suggested that 
the mass of the first stars is limited by the declining accretion rate 
near the $\Omega\Gamma$-limit, while we here showed that 
the accretion continues even after the star reaches the $\Omega\Gamma$-limit.
In this section, we compare our result with that of the previous works.

PN91 calculated the steady accretion-disk solutions assuming the polytropic 
gas with index $\gamma=2$ for given sets of 
stellar rotation rate 
$\Omega_*$ and angular momentum flux $j$ onto it.
In their solutions, the disks have increasingly large scale height 
toward the inner edge. 
The authors identified it as the transition from the disk to 
the stellar surface and defined the star-disk boundary by the condition $H/r=0.1$.
Calculating the structure from the disk 
to the stellar surface continuously, they could obtain the 
$\Omega_*-j$ relation by this additional condition.

On the other hand, if the stellar radiation effect is included, 
such a high scale-height region, which can be regarded as the stellar surface, 
does not always appear \cite[]{1996ApJ...467..749P}.
We thus cannot use the condition such as $H/r=0.1$ to identify the star-disk boundary. 
(cf. the disk scale height for the $5M_{\odot}$ case in Figure \ref{fig:evolv}).
In fact, such a condition on the aspect ratio $H/r$ is not adopted to identify the star-disk boundary
in \cite{1996ApJ...467..749P}, who studied the accretion disks around FU Ori-type stars 
using the method similar to ours.
\cite{1996ApJ...467..749P} instead used a condition on the radial velocity 
to identify the stellar surface, but its validity is uncertain for  
obtaining the $\Omega_*-j$ relation. 
In this work, we speculated that 
the disk solution with $j=0$ is realized after the star reaches the $\Omega\Gamma$-limit
as long as such a solution exists.
Although beyond the scope of this paper, 
we point out that the $\Omega_*-j$ relation can be obtained 
in principle
by solving the detailed star-disk structures consistently
with an appropriate boundary condition between them.

\subsection{Radiative feedback from the inner disk to the star}
\label{Teff}

As seen in Section \ref{result}, the temperature in the inner disk sometimes 
exceeds the stellar effective temperature.
If the star is covered by such hot matter around the equatorial zone, 
the radiation hardly diffuses out in this direction,
but is mostly channeled in the polar direction.
If so, the radiation flux from the star to the disk 
would be smaller than the bare stellar photospheric value used in this work.
We here estimate the temperature at the inner disk edge to assess this effect. 
For simplicity, we assume the Keplerian rotation for the disk and 
conservatively neglect radiative heating by the star. 
In this case, the temperature at the inner edge is estimated as
\begin{eqnarray}
 T&=&4.3\times 10^5 [{\rm K}] \left(\frac{\alpha}{10^{-2}}\right)^{1/5}
\left(\frac{\dot{M}}{4\times 10^{-3}\ M_{\odot}\ {\rm yr^{-1}}}\right)^{2/5} \nonumber \\
&&\times\left(\frac{M_*}{20M_{\odot}}\right)^{3/10}
\left(\frac{r_{\rm in}}{0.1 \ {\rm au}}\right)^{-9/10},
\end{eqnarray}
where we have used the electron scattering value $\kappa=0.35\ {\rm cm^2\ g^{-1}}$ for the opacity assuming the complete ionization.
This temperature is much higher than the typical stellar surface value of a few times $10^4$ K.
In this case, the radiation is inward rather than outward around the star-disk edge, which   
would reduce radiation force exerted by the star.
Such effect should be studied in future by simultaneous modeling of the star and disk.

\section{Summary}
\label{summary}
We have considered whether a rotating primordial protostar can 
continue accreting material from the circumstellar disk even after the star reaches 
the critical configuration where the sum of 
the outward centrifugal and radiation forces balances the inward pull of gravity at the surface, 
or so-called the $\Omega\Gamma$-limit. 
For further accretion, the gas should not bring angular momentum to the star.
Once the protostellar radiation becomes intense and the radiation force becomes 
comparable to the gravity as well as the centrifugal force, 
it significantly affects the structure of accretion disk.
We have calculated a series of the steady accretion disk solutions by 
taking into account the effect of the radiation force/heating by the protostar.
The main results can be summarized as follows:

\begin{itemize} 
 \setlength{\itemindent}{0pt}
 \setlength{\leftskip}{10pt}
\item 
The star can accrete the gas from the disk without receiving 
the angular momentum even at the rotation rate as low as $\Omega_*/\Omega_{K*}\simeq 0.1$ 
(or lower), in contrast with PN91's analysis without the radiation effect,
where such accretion was possible only for $\Omega_*/\Omega_{K*}\gtrsim0.9$. 

\item
The circumstellar disk without net angular momentum influx to a star at the $\Omega\Gamma$-limit has following characteristics.
Around the inner edge, outward radiation force contributes non-negligibly in the 
force balance, where the sum of radiation and centrifugal forces is approximately equal to the inward pull of gravity. 
In terms of the energy, heating due to the diffusing stellar radiation balances with 
the radiative cooling from the disk surfaces with little contribution by the viscous heating.
For the protostars with $M_{\ast} \gtrsim50M_{\odot}$
and rotating at $\Omega_{\rm crit}\lesssim 0.4\Omega_{\rm K*}$, the inner edge of the disk bloats due to high temperature and surface density;  
an envelope-like structure emerges
as a result of accumulation of a large amount of gas around the inner edge, which is caused by the high mass accretion rate with no net the angular momentum influx. 

\item
Since the protostar does not receive the angular momentum by such a mode of accretion, 
the mass growth is not hindered by the $\Omega\Gamma$-limit.
For the protostars rotating at the critical rate for the $\Omega\Gamma$-limit, 
we have shown that it can grow at least up to $M_*=70M_{\odot}$ (and probably more) 
in the case of the accretion rate ${\dot M}=4\times 10^{-3} M_{\odot}\ {\rm yr^{-1}}$.

 \end{itemize}

Here, we have calculated the disk solutions 
for arbitrary combinations of the stellar rotation rates and the angular momentum influx to the star. 
In reality, however, there is some relation among those paramenters 
and the latter should be uniquely determined 
by the condition of the disk and star system. 
We defer pinpointing such relation by consistently 
calculating the protostar and disk structures to future works.

\section*{Acknowledgments}
We thank Takashi Hosokawa for providing the protostellar evolution data and for valuable comments.
We also thank Kazu Sugimura, Shigeo Kimura, and Hajime Fukushima for fruitful discussion.
This work is supported in part 
by NAOJ ALMA Scientific Research Grant Numbers 2016-02A (SZT) and
by MEXT/JSPS KAKENHI Grant Number 25287040, 17H01102 (KO).

\appendix
\section{Dependence of the disk structures on the protostellar spin and the angular momentum influx}
\label{Omega_Jdot}
In Section \ref{evolv}, we investigate the time evolution of the disk structure with $\Omega_*=\Omega_{\rm crit}$ and $j=0$.
In this section, we investigate the disk structures for different $\Omega_*$ and $j$.

\subsection{Dependence on $\Omega_*$}
\label{Omega_dependence}
Disk structures for $\Omega_*<\Omega_{\rm crit}$ and $\Omega_*>\Omega_{\rm crit}$ are very different.
In the former case, the gravity is larger than the sum of the radiation and centrifugal forces, while 
the opposite is true in the latter case.  
In this section, we show examples for the case of $M_*=35M_{\odot}$, where the critical angular velocity 
is $\Omega_{\rm crit}/\Omega_{\rm K*}\approx 0.7$ and the radiation force is about a half of the gravitational force.

\begin{figure}
 \begin{center}
  \includegraphics[width=8cm]{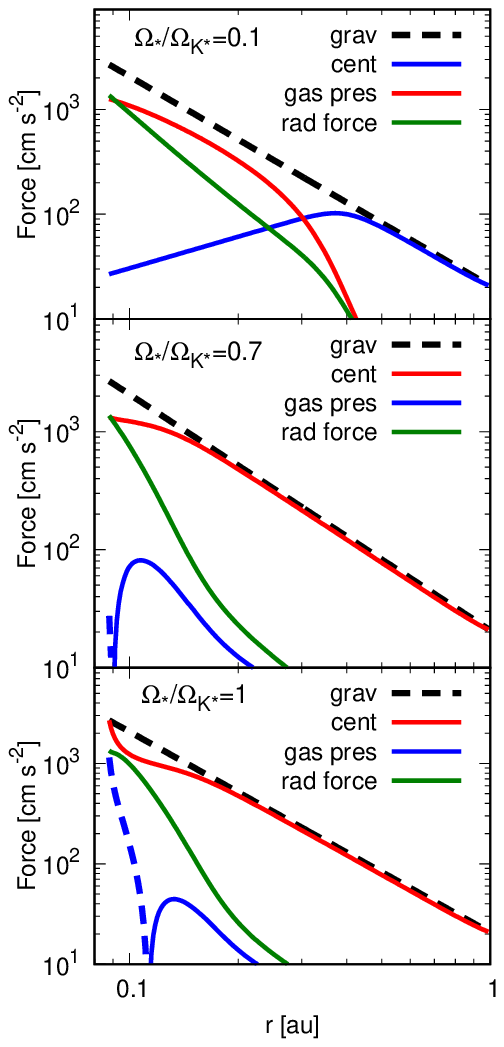}
  \caption{
Radial distribution of the force components in the disk 
for $\Omega_*/\Omega_{\rm K*}=0.1$, 0.7, and 1 from top to bottom.
The dashed and solid lines show the inward and outward forces, respectively.
The protostar mass is 35$M_{\odot}$, which corresponds to $\Omega_{\rm crit}/\Omega_{\rm K*}=0.7$.
The forces are balanced in the disk, and the gas pressure gradient 
acts outward (inward) for $\Omega_*/\Omega_{\rm K*}=0.1$ ($\Omega_*/\Omega_{\rm K*}=1$, respectively).
}\label{fig:force_spin}
 \end{center}
\end{figure}

\begin{figure}
 \begin{center}
  \includegraphics[width=8.2cm]{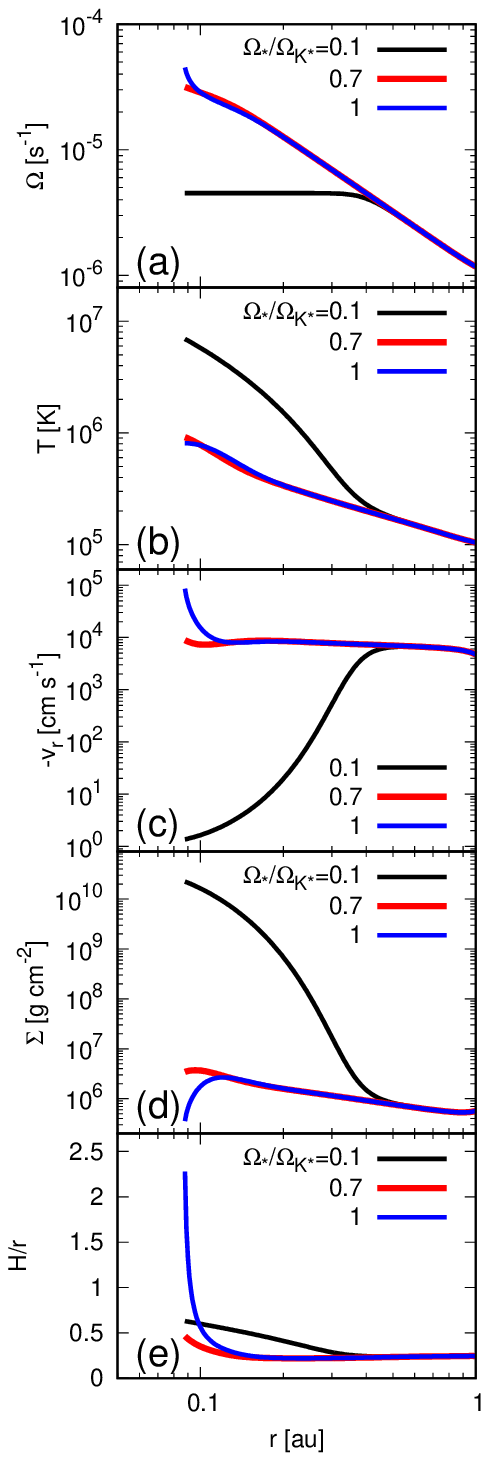}
  \caption{
Comparison of the disk structures for the three cases shown in Figure \ref{fig:force_spin}. 
The five panels from top to bottom show  (a) the angular velocity, (b) the temperature, (c) the radial velocity, (d) the surface density, 
and (e) the aspect ratio as functions of the radius.
}\label{fig:disk_st_spin}
 \end{center}
\end{figure}

Figure \ref{fig:force_spin} shows the centrifugal, radiation, and gravitational forces 
for stars with angular velocity $\Omega_*/\Omega_{\rm K*}=0.1,\ 0.7,\ {\rm and }\ 1$ from top to bottom.
For $\Omega_*/\Omega_{\rm K*}=0.1$ the centrifugal force is negligible around the inner edge of the disk.
Instead, the gas pressure supports the disk against the gravitational pull.
On the other hand, in the case of $\Omega_*/\Omega_{\rm K*}=1$, the inward pressure gradient force balances with the outward radiation force.
Figure \ref{fig:disk_st_spin} shows the angular velocity, temperature, radial velocity, surface density, and aspect ratio 
for the same cases.
The disk structure of $\Omega_*/\Omega_{\rm K*}=0.1$ is similar to that of 
$\Omega_*=\Omega_{\rm crit} \ll \Omega$ (cf. Figure \ref{fig:evolv} $M=70M_{\odot}$) in the sense that
the angular velocity is almost constant ($\Omega\sim 0.1\Omega_{\rm K*}$) in the inner wide region,
and the infall velocity is small (Figure   \ref{fig:disk_st_spin} a and c).
The resultant high surface density and the temperature cause large pressure gradient force.
On the other hand, the disk structure of $\Omega_*/\Omega_{\rm K*}=1$ is similar 
to that of $\Omega_*/\Omega_{\rm K*}=0.7$, except that $\Omega$ becomes extremely large at the inner edge.
Due to the large infall velocity here by $v_r \propto d\Omega/dr$ (Equation (\ref{eq:dOmegadr})), 
the surface density becomes low in this region (Equation (\ref{eq:Sigma})).
As a result, the pressure gradient force now acts inward, which allows inward accretion flow 
even though the sum of the centrifugal and radiation forces exceeds the gravity.

\subsection{Dependence on $j$}
\label{j_dependence}
\begin{figure}
 \begin{center}
  \includegraphics[width=8.5cm]{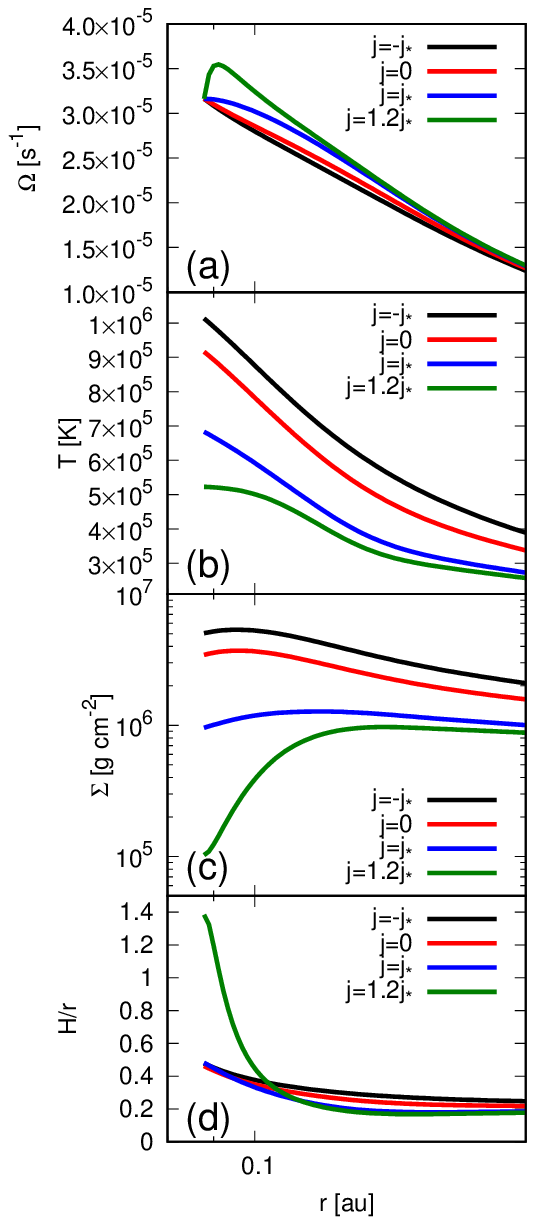}
  \caption{
Comparison of the disk structures for $j=-j_*,\ 0,\ j_*,\ {\rm and }\ 1.2j_*$.
The four panels from top to bottom show (a) the angular velocity, (b) the temperature, 
(c) the surface density, and (d) the aspect ratio.
The protostar mass and rotation rate are $M=35M_{\odot}$ and $\Omega_*=\Omega_{\rm crit}$, respectively.
}
  \label{fig:disk_str_j}
 \end{center}
\end{figure}

The disk structure depends also on the parameter $j$, which 
corresponds to the specific angular momentum flux 
if the angular momentum transfer is only by the advection (see Equation \ref{eq:Jconserv}).
In the case of $j<0$ ($j>0$), outward angular momentum transport is larger (smaller) 
than the angular momentum advection, and so the angular momentum of the protostar decreases 
(increases, respectively).
Figure \ref{fig:disk_str_j} shows the disk structures for 
$j=-j_*,\ 0,\ j_*,\ {\rm and }\ 1.2j_*$, where $j_*=\Omega_*R_*^2$.
Here, the protostar mass is $M=35M_{\odot}$ as in the previous section, and the angular velocity of the protostar is fixed to $\Omega_*=\Omega_{\rm crit}\ (=0.7\Omega_{\rm K*})$.
The disk structure strongly depends on whether $j$ is larger than the specific angular momentum 
of the gas $\Omega r^2$ because $j$ appears as a form of $\Omega-j/r^2$ in the Equation (\ref{eq:dOmegadr}).
As a result, only the structure near the inner edge is altered by different $j$ 
since $\OK \propto r^{-1.5}$ while $j/r^2\propto r^{-2}$.

In cases with $j=-j_*$ or 0, the angular momentum is always transferred outward by the torque.
In the former case, the angular momentum transfer by the torque is twice that by the advection 
at the inner edge of the disk, while they are the same in the latter.
As shown in Figure \ref{fig:disk_str_j}, the disk structures for $j=-j_*\ {\rm and}\ 0$ 
are qualitatively the same: 
$\Omega$ decreases monotonically outward and their differences in the temperature, 
the surface density, and the aspect ratio are within a factor of two.
In the case of $j=j_*$, no angular momentum is extracted from the star by the torque
and the $\Omega$ distribution is flat near the inner edge.
The disk structure for $j=1.2j_*$ is very different from the other cases.
With $j>j_*$, the angular momentum is transported inward by the torque around the inner edge, 
which requires $d\Omega/dr>0$.
The infall velocity increases with $d\Omega/dr$ 
as 
\begin{equation}
 v_r=-\frac{\nu}{j/r^2-\Omega}\frac{d\Omega}{dr}
\end{equation}
from Equation (\ref{eq:dOmegadr}), 
and the surface density $\Sigma \propto v_r^{-1}$ (Equation (\ref{eq:Sigma})) becomes 
small in the region.
As the surface density decreases, 
the scale height increases drastically 
by the relation $H=2aT^4/(3\Sigma\Omega_*^2)$ which is valid when the radiation pressure is dominant 
(see Equation (\ref{eq:H})).
 
Outside $\sim 0.1$ au, no significant difference is seen in the distributions 
of the angular velocity and the aspect ratio. 
Here, the surface density and the temperature are larger for smaller $j$. 
Since more angular momentum need to be transported 
from the star to the disk in lower $j$ solution, 
the disk material cannot accrete swiftly 
onto the star and get stuck in the disk.  
\end{document}